\documentclass[%
 reprint,
 amsmath,amssymb,
 aps,
]{revtex4-2}

\usepackage{graphicx}
\usepackage{dcolumn}
\usepackage{bm}%

\usepackage[utf8]{inputenc}
\usepackage[english]{babel}
\usepackage[T1]{fontenc}
\usepackage{amsmath}
\usepackage{amsfonts}
\usepackage{tikz}
\usepackage{lipsum}
\usepackage{braket}
\usepackage{comment}
\usepackage{soul}
\usepackage[normalem]{ulem}
\usepackage{xpatch}
\usepackage{csquotes}

\begin{document}

\preprint{APS/123-QED}

\title{Hybrid quantum-classical algorithm for transverse-field Ising model in the thermodynamic limit}

\author{Sumeet}
\email{sumeet.sumeet@fau.de}
\author{M. H\"ormann}
\author{K. P. Schmidt}
 
\affiliation{%
 Department of Physics, Friedrich-Alexander-Universit\"at Erlangen-N\"urnberg (FAU), 91058 Erlangen, Germany}%




\date{\today}

\begin{abstract}
We describe a hybrid quantum-classical approach to treat quantum many-body systems in the thermodynamic limit. This is done by combining numerical linked-cluster expansions (NLCE) with the variational quantum eigensolver (VQE). Here, the VQE algorithm is used  as a cluster solver within the NLCE. We test our hybrid quantum-classical algorithm (NLCE$+$VQE) for the ferromagnetic transverse-field Ising model on the one-dimensional chain and the two-dimensional square lattice. The calculation of ground-state energies on each open cluster demands a modified Hamiltonian variational ansatz for the VQE. One major finding is convergence of NLCE$+$VQE to the conventional NLCE result in the thermodynamic limit when at least $N/2$ circuit repetitions (known as layers) are used in the VQE ansatz for each cluster with $N$ sites. Our approach demonstrates the fruitful connection of techniques known from correlated quantum many-body systems with hybrid algorithms explored on existing quantum-computing devices. 

\end{abstract}

\maketitle

\section{Introduction}

Quantum computers in the noisy intermediate scale quantum (NISQ) era are essential tools to test the potential advantage over conventional classical methods for solving quantum many-body systems. Therefore, it is inevitable to explore the advantages that quantum algorithms can impart to the existing problems of classical computers, which are limited by the exponential growth of Hilbert space with the increase in size of quantum systems. Indeed, all existing classical methods for solving quantum many-body lattice systems such as tensor networks ~\cite{Verstraete2006, Hastings2007}, quantum Monte Carlo techniques ~\cite{Whitlock2006}, density matrix renormalization group ~\cite{Stlund1995,Vidal2007}, etc., have their individual limitations. As a consequence, hybrid quantum-classical algorithms such as variational quantum eigensolver (VQE)\cite{Peruzzo2014}, quantum adiabatic optimization algorithm (QAOA)\cite{Wang2022}, CQE (contracted quantum eigensolver) \cite{Smart2021}, etc., are gaining popularity by promising to solve the problem of exponential Hilbert space growth with a potential speed-up and within a range of acceptable errors. 

The hybrid variational quantum eigensolver (VQE) algorithm, developed by Perruzo
\textit{et al.} \cite{Peruzzo2014}, uses a union of classical and quantum computation to minimize a cost function.  The VQE algorithm is not only more resilient to quantum errors as compared to other quantum algorithms \cite{Tilly2022} but also requires low coherence times \cite{McClean2016}, making it one of the most promising NISQ algorithms. Additionally, VQE can be implemented on all quantum hardware. It can therefore be an efficient tool to investigate the properties of quantum many-body systems.

However, the existing NISQ devices and the use of quantum-classical algorithms like VQE are highly limited due to small number of available (noisy) qubits. Obviously, it would be desirable to overcome this limitation and to determine, at least in an approximate fashion, the properties of quantum many-body lattice problems in the thermodynamic limit exploiting available quantum computers. Interestingly, in classical computation, several approaches exist that can be used to approximate the properties of a quantum system in the thermodynamic limit. In particular, series expansion methods such as linked-cluster expansions (LCE) \cite{Sykes2004, Wales2006} exploit a full graph decomposition to determine physical quantities in the thermodynamic limit by perturbative calculations on finite graphs. LCEs were first used for perturbative calculations of the ground-state energy of quantum many-body systems and then generalized to calculate excitation energies \cite{Gelfand1996}. A non-perturbative extension of LCEs are numerical linked-cluster expansions (NLCE) \cite{Rigol2006, Tang2013}, which deduce energies and observables in the thermodynamic limit by replacing perturbation theory on graphs with non-perturbative tools like exact diagonalization (ED). This concept was applied to non-perturbative calculations of thermodynamic quantities \cite{Rigol2006,Rigol2007} and for the calculation of ground-state properties of quantum many body systems \cite{Yang2011,Khatami2011a,Ixert2016}, albeit the idea already had been proposed much earlier \cite{Irving1984}. It has further been used successfully to study properties of various materials \cite{Rigol2007a, Khatami2011, Khatami2012, Applegate2012}, quantum dynamics \cite{Guardado-Sanchez2018, Mallayya2018}, thermodynamic properties of systems \cite{Tang2015, Mulanix2019}, and was applied to optical lattice experiments with fermions \cite{Khatami2011a,Khatami2011,Khatami2012b,Hart2015}. {\mbox NLCEs} were also successfully applied to calculate excitation energies of quantum many-body systems using contractor renormalization group (CORE) \cite{Morningstar1996} with ED and graph-based continuous-unitary transformations (gCUT) \cite{Yang2011,Coester2015}. Only recently the approach utilizing ED was generalized to non-parity broken models and arbitrary number of excitations using projective cluster-additive transformations (pCAT) \cite{Hormann2023}. For LCE calculations, convergence is limited to the convergence radius of the perturbative expansion. The situation is less clear for NLCEs. One expects quantities to converge as long as the subspace of interest is gapped. This is indeed given for our application of \mbox{NLCEs} to the ground-state energy of the TFIM. NLCEs are then known to converge when the spatial correlations in the system do not exceed the sizes of the graphs considered. However, the use of ED as solver on the graphs limits current NLCE calculations due to the exponential growth of the Hilbert space. In addition, the number of graphs increases exponentially or worse with the number of sites.

In this work we explore a hybridization of NLCE and the quantum-classical algorithm VQE, dubbed NLCE+VQE, in order to go beyond their respective limitations and to fuse their advantages. Hence, we integrate VQE as a graph solver within the NLCE replacing ED. 
For this purpose, we use ideal statevector simulations, i.e. we conduct simulations using statevectors avoiding any measurement errors and quantum noise for the VQE throughout the paper.  

For all current VQE applications on existing quantum computing devices a true quantum advantage is hard or even impossible to achieve due to the amount of noise. The same is true for NLCE+VQE. We expect that the situation will gradually improve in the near future when errors in the quantum hardware are more and more suppressed. However, a true breakthrough will be the implementation of quantum error correction and the realization of fault tolerant quantum computers. Then, assuming a working optimization scheme, NLCE+VQE can be used to perform NLCEs up to cluster sizes that are not feasible to reach with conventional NLCE+ED. 
In this article we want to demonstrate that at least up to system sizes of conventional NLCE+ED it is possible to obtain quantititative results with NLCE+VQE. This does not prove that scaling up to even larger system sizes will work, but is an essential requirement. To this end, we identify challenges and develop strategies related to finding the global minimum in the optimization landscape within VQE. In addition, we investigate how many layers in the VQE ansatz are needed for quantitative results. This is important because a stable and working optimization procedure and a polynomial scaling of layers are both necessary for applying NLCE+VQE with true quantum utility beyond NLCE+ED. 
We conduct simulations using statevectors avoiding any measurement errors and quantum noise for the VQE throughout the paper. Typically, we use Cirq \cite{cirq_developers_2023_10247207} for the simulations.
We perform exact diagonalization using the QuSpin package ~\cite{Lehoucq2013} to benchmark our results in NLCE+ED. The calculation of ground-state energies on each open cluster demands a modified Hamiltonian variational ansatz for the VQE. We benchmark NLCE+VQE for the ferromagnetic transverse-field Ising model (TFIM) by calculating the ground-state energy of the high-field polarized phase in a quantitative fashion in the thermodynamic limit. The TFIM is introduced in section~\ref{sec:model}. Our NLCE+VQE approach as well as details for NLCE and VQE are described in section~\ref{sec:nlcevqe}. We use VQE algorithm to compute the ground-state energy of finite clusters of a lattice model and then embed them on the infinite lattice using the NLCE. In order to improve the optimization process for the NLCE$+$VQE method, we also study the ground-state energy of periodic TFIM clusters in section~\ref{sec:periodic}. These findings are used to improve the ansatz and initial guess parameters for implementing the VQE algorithm to graphs in section~\ref{sec:cluster-solver}. We provide NLCE+VQE results for the TFIM on the chain and the square lattice in section~\ref{sec:calculation thermodynamic limit} together with an analysis of the shot-noise error. Finally, we conclude this work in section~\ref{sec:conclusion}.

\section{Transverse-field Ising model} \label{sec:model}

The transverse-field Ising model is one of the paradigmatic models of quantum many-body physics. Since it is exactly solvable in one dimension, the TFIM can act as a good system to benchmark the efficiency of quantum-classical hybrid algorithms against classical computational approaches. The Hamiltonian for the transverse-field Ising model is given by 
\begin{equation}\label{ham_tfim}
    H = - \,h \sum_{\nu} Z_\nu - J  \sum_{<\nu,\,\mu>}X_\nu X_\mu,
\end{equation}
where $J$ is the coupling constant and $h$ is the external field applied in a transverse direction.  $Z$ and $X$ are Pauli matrices. For $J\gg h$ and $J>0$ ($J<0$), the model is in a ferromagnetic (antiferromagnetic) Ising-phase when we restrict to unfrustrated lattices. The Hamiltonian given in Eq.~\ref{ham_tfim} has $\mathbb{Z}_2$ symmetry and remains invariant under the action of flipping all the spins. The one-dimensional transverse-field Ising chain is exactly solvable by mapping the Hamiltonian to a quadratic fermionic one with the Jordan-Wigner transform and subsequently solving it with the Bogoliubov transformation \cite{Pfeuty1970}. The system is in a disordered phase for $J/h < 1$ and in the ordered Ising phase for $J/h > 1$.  The quantum phase transition occurs at $J/h=1$ and has same universality class as the two-dimensional classical Ising model. In general, one can map the d-dimensional TFIM to a $d+1$-dimensional classical Ising model. For the two-dimensional square lattice, the TFIM undergoes a quantum phase transition at $J/h\approx 0.328$ in the 3d Ising universality class \cite{He1990, Hesselmann2016}. While on bipartite lattices, the ferromagnetic and antiferromagnetic TFIM behave identically, on non-bipartite lattices, the behaviour of the antiferromagnetic TFIM can be very different. For example on the frustrated triangular lattice, it undergoes a phase transition in the 3dXY-universality class and on the frustrated Kagome lattice, the model is in a disordered phase for any $h>0$ \cite{Powalski2013, Moessner2001}. In this work, we restrict ourselves to the non-frustrated TFIM on a chain and square-lattice geometry and calculate the ground-state energy in the disordered phase with the NLCE+VQE approach.\\

\section{NLCE$+$VQE}\label{sec:nlcevqe}

\subsection{Numerical linked-cluster expansions}\label{sec:NLCE}

NLCEs use calculations on finite clusters to approximate properties of a system in the thermodynamic limit. This method exploits that some quantities are additive, which we will explain for the ground-state energy in the following. We refer the reader to chapter 4 of \cite{oitmaa2006series} and \cite{Tang2013} for pedagogical introductions to the framework of  perturbative and numerical linked-cluster expansions.

Any ground-state expectation value $P\equiv \Braket{\mathcal{P}}$ of a local observable $\mathcal{P}$ that allows for a decomposition $\mathcal{P} = \mathcal{P}_A\otimes 1_B + 1_A \otimes \mathcal{P}_B$ on two disconnected clusters $A$ and $B$, where $\mathcal{P}_{A,B}$ only acts on the space $A,B$ and the local observables fulfill $[P_A\otimes 1_B, 1_A \otimes P_B]=0$, is 
\begin{equation}
    P=P_A + P_B\, .
    \label{eq::additivity ground-state energy}
\end{equation}
This follows from the ground-state being a product state on $A$ and $B$, $\ket{\Psi}_{A\cup B} = \ket{\Psi}_A \otimes \ket{\Psi}_B$, and
\begin{equation}
    \mathcal{P}_{A}\otimes 1_B \ket{\Psi}_A \otimes \ket{\Psi}_B = \mathcal{P}_A \ket{\Psi}_A \otimes \ket{\Psi}_B = P_A \ket{\Psi}_A \otimes \ket{\Psi}_B.
\end{equation} We now recursively define reduced quantities by 
\begin{equation}
    \bar{P}_c = P_c - \sum_{c^\prime \subset c } \bar{P}_{c^\prime}\,.
    \label{eq::reduced energies}
\end{equation}
From Eq.~\eqref{eq::additivity ground-state energy} and Eq.~\eqref{eq::reduced energies} it follows that $\bar{P}_c=0$ for a non-linked cluster $c$ since it is equivalent to $P_c = \sum_{c^\prime \subseteq c} \bar{P}_{c^\prime}$. As all other reduced contributions are zero, one can perform a linked-cluster expansion for the value of $P$ per site, 
\begin{equation}\label{eq:nlce-energy}
    p = \sum_c w_c \bar{P}_c\, .
\end{equation}
The sum runs over all linked clusters $c$ and the embedding factor $w_c$ is determined by the numbers of embeddings for this graph on the lattice. This follows from Eq.~\eqref{eq::reduced energies} applied to the lattice and the assumption of invariance of all topologically equivalent subgraphs of the lattice. One such local observable is in our case the Hamiltonian itself and $P$ equals the energy $E$. The reduced energy is $e$. This is what we will calculate throughout the paper. One usually has to truncate the expansion at some point. We are investigating disordered phases with finite correlation lengths. It is then straightforward to truncate in terms of the size of the largest cluster used in the NLCE.

\begin{figure}
    \centering
    \includegraphics[width=80mm]{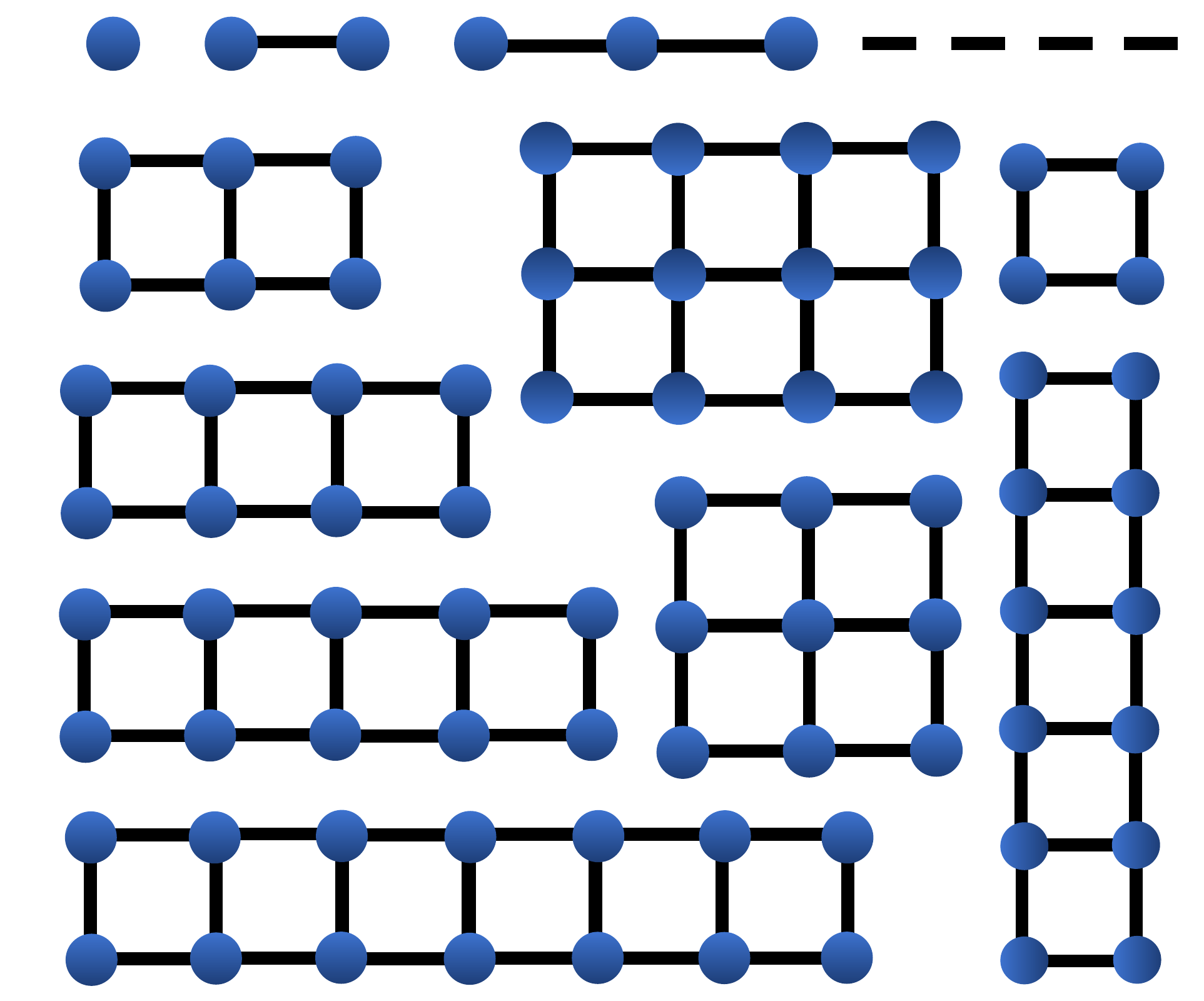}
    \caption{This figure shows the rectangular graphs used in the rectangular-graph expansion of NLCE for the TFIM on the chain (first line) and the square lattice (all graphs) up to the largest cluster size of 14 spins. The dashed line indicates the missing chain segments.} 
    \label{nlce_rect_fig}
\end{figure}

The decomposition of a lattice into finite clusters used for NLCEs is not unique. Often \mbox{NLCEs} are performed with all graphs of the lattice, also called full-graph expansion. In this work, we use a rectangular graph expansion \cite{Enting1996} for the chain and the square lattice (see Fig.~\ref{nlce_rect_fig}). It coincides with the full graph expansion on the chain. Rectangular graph expansions have been used in various LCE and NLCE applications before \cite{Dusuel2010,Kallin2013a,Stoudenmire2014a,Ixert2016,Richter2020,Gan2020}. In one dimension, $e = \sum_c w_c \bar{E}_c$ reduces to $e = E_L-E_{L-1}$ for all graphs up to length $L$. In two dimensions this does not hold any more because the embedding factors are not the same for all graphs and we need all reduced energies of rectangles up to a certain size to calculate the ground-state energy. The number of graphs in the rectangular-graph expansion scales polynomially with the number of sites. Fig.~\ref{nlce_rect_fig} shows the graphs used in rectangular-graph expansion up to 14 spins which includes one spin, one-dimensional chains from one to fourteen spins, ladders, and square graphs. The order for this expansion is defined by the number of sites $N =L_X \times L_Y$, where $L_X$ and $L_Y$ are the widths and heights of the rectangles.

\subsection{The variational quantum eigensolver algorithm}

The ground-state energy of the TFIM can be computed using the Rayleigh-Ritz variational principle, which states that a parametrized wavefunction can be used to determine the upper bound to the ground-state energy 
\begin{equation}
    E_{\mathrm{gs}} \leq E(\theta) = \langle \psi(\theta)|H|\psi(\theta) \rangle.
\end{equation}
This principle serves as the basis for the variational quantum eigensolver (VQE) algorithm \cite{Ritz1909}. VQE is a quantum-classical hybrid algorithm that uses a quantum computer to determine the eigenstates of a quantum many-body Hamiltonian and  a classical computer to implement an optimization routine to minimize a cost function, which in our case, is the energy. A parameterized wavefunction is prepared using a unitary operation $U(\theta)|\psi\rangle$ in the first step called state preparation. In the second step, the expectation values $\langle \psi(\theta)|H_s| \psi(\theta) \rangle$ are determined on a quantum machine, where $H_s$ are the different terms in the Hamiltonian. The expectation values are further added to determine the total energy eigenvalue, which is then sent to a classical optimizer. The optimizer minimizes the energy in an iterative loop which goes over the first step, i.e. state preparation with new parameters decided by the optimization method. This is repeated until a global minimum for energy is obtained, hence, resulting in the ground-state energy.\\

All variational quantum algorithms including VQE suffer from a common issue of difficult parameter optimization. One often finds non-convex energy landscapes in the ground-state energy of quantum many-body systems due to which the VQE algorithm is highly susceptible to errors caused by barren plateaus \cite{Wecker2015, Cerezo2021} and local minima \cite{Bittel2021} during optimization. Therefore, the design of the variational ansatz is of core importance to the VQE algorithm. VQE can promise a potential advantage over conventional classical methods of computation if the quantum circuit for the ansatz can represent sufficiently accurate trial wavefunctions. \\

The optimization strategy used in VQE is extremely important. VQE optimization is shown to be NP-hard \cite{Bittel2021}. A good reference state or well chosen initial guess parameters assist the VQE optimization to reach the global minima for the cost function. We investigate how to find these initial guess parameters in the subsequent sections.\\

 \subsubsection{Ansatz}

 The choice for $U(\theta)$ is very important for the VQE algorithm. The ansatz must have the ability to span a substantial number of states in the Hilbert space, a property known as expressibility \cite{Nakaji2021, Sim2019}. It denotes the accuracy with which an ansatz in VQE can determine important low-energy states when the optimization reaches a global minimum. Another salient feature of the ansatz is the trainability that describes the ability of the ansatz to be optimized on a quantum device. This includes the shape of optimization landscape, optimum number of parameters and susceptibility to get stuck on barren plateaus or in local minima. The ansatz apart from being sufficiently expressible must contain an appropriate number of parameters so that it can approximate the ground-state wavefunction with acceptable accuracy. However, it should not be excessively expressible so that the wavefunction of interest becomes intractable. Also local minima are more likely to occur, if it is too expressible. Apart from these aspects, another relevant property for the ansatz is the scaling and complexity of quantum circuits with system size. It is important to have an optimum circuit depth for the ansatz that does not scale exponentially with the system size.\\
 
 One good ansatz that calibrates the aforementioned features is the Hamiltonian variational ansatz (HVA) \cite{Wecker2015, Wiersema2020}, which is inspired by the quantum approximate optimization algorithm (QAOA) \cite{Wang2022} and adiabatic quantum computation \cite{Farhi2000}. HVA has been used for various models such as the transverse field Ising model, XXZ model, Hubbard model  and many others \cite{Wecker2015,Wiersema2020, Wang2022a, Wiersema2020, Park2023, Mele2022, Dagotto1985, AnselmeMartin2022}. For the Hamiltonian of the form $H = \sum_s H_s $, where the summation runs over all the terms in the Hamiltonian, the HVA is given by 
 \begin{equation}
     U(\Theta) = \prod_l \big(\prod_s e^{\mathrm{i}\theta_{s,l} H_s} \big),
     \label{eq::HVA_basic}
 \end{equation}
where $l$ denotes to the number of layers of HVA. There is still freedom to choose the ordering of operators $H_s$ in Eq.~\eqref{eq::HVA_basic}. For some classes of Hamiltonians one can write $H=H_A+H_B$ with local terms in $H_A$ or $H_B$ commuting, but $[H_A,H_B]\neq 0$. It is then natural to choose an ansatz, where first a layer of local (potentially differently weighted) terms of $H_A$ is put in the circuit, followed by a layer of $H_B$ and so on \cite{Wang2022}.\\

\begin{figure}
     \centering
     \includegraphics[width=80mm]{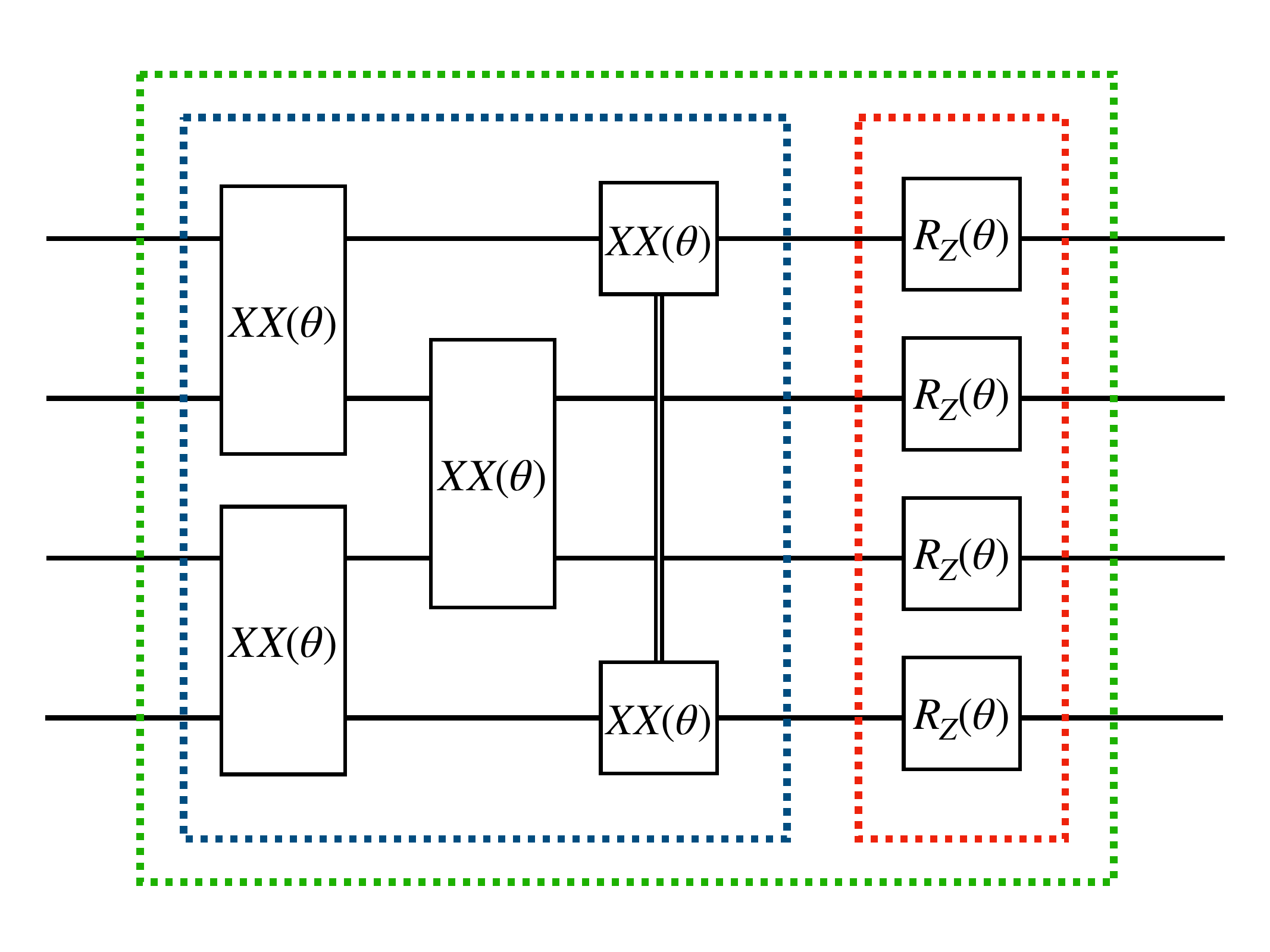}
     \caption{Quantum circuit for the Hamiltonian variational ansatz for the transverse-field Ising model with periodic boundary conditions. The horizontal lines represent the qubits and the vertical boxes represent quantum gates. $e^{\mathrm{i}\theta \sigma_x \sigma_x}$ term in the ansatz is implemented using the $XX(\theta)$ entangling quantum gate. The quantum gate $R_z(\theta)$ implements the Euler rotation along Z axis.}
     \label{fig:HVA}
 \end{figure}

The quantum circuit corresponding to the HVA is shown in Fig.~\ref{fig:HVA} for the TFIM. The first block in blue color implements the Ising part of the Hamiltonian $e^{\mathrm{i}\sum_j\theta^{XX}_{\nu,\mu,l}X_\nu X_{\mu}}$ along with the boundary term. The second red block implements the transverse field term of the Hamiltonian, i.e. $e^{\mathrm{i}\sum_\nu\theta^{Z}_{\mu,l} Z_j}$. Both these blocks make up one layer, represented by a green block. These layers in the ansatz are repeated while calibrating expressibility and entanglement in the quantum circuit. An interesting feature of HVA is that its structure does not require entanglement between more than two qubits for models with only two qubit-interactions, which makes it hardware friendly. A convenient reference state can be chosen as
\begin{equation}
    \ket{\mathrm{ref}}\equiv \ket{\uparrow\dots \uparrow},
\end{equation} 
i.e. the fully polarized state. The unitary circuit is applied to this reference state and the energy of $U(\Theta)\ket{\mathrm{ref}}=\ket{\Psi(\Theta)}$ is measured. More layers impart more parameters for optimization which may or may not help reaching the global minima depending on the optimization procedure. The energy is expected to reduce with each additional layer as $E_{l+1}(\theta)< E_{l}(\theta)$ until a convergence is reached. We investigate this convergence in subsection \ref{sec:periodic} and section \ref{sec:calculation thermodynamic limit}, where we will see that the number of layers needed depends on the system size.

\subsubsection{Optimization}\label{optimization}

The VQE algorithm is an optimization problem at heart as it heuristically constructs an approximation of the wavefunction through iterative training of parameters in the ansatz. A good optimization strategy helps to reach a well-approximated solution to the minimization procedure within an acceptable number of iterations. Therefore, starting with parameters already close to the real solution makes it easier for the optimizer to find global minima and also reduces the number of iterations required for optimization. Apart from that, restricting the parameter space for the optimization process decreases the potential of getting stuck on barren plateaus and local minima. Barren plateaus are a result of occurrence of vanishing gradients in the gradient descent method.\\

The optimization parameters can be constrained in the range $[0,\pi]$ since the HVA ansatz for the TFIM is periodic in $\pi$. This follows from
\begin{equation}
\begin{aligned}
 e^{\mathrm{i}(\varphi+\pi) X_\nu X_\mu} & = e^{\mathrm{i}(\varphi+\pi)}e^{\mathrm{i}(\varphi+\pi) (X_\nu X_\mu-1)} \\ 
 & = - e^{\mathrm{i}\pi (X_\nu X_\mu-1)} e^{\mathrm{i}\varphi X_\nu X_\mu} \\
 & = - e^{\mathrm{i}\varphi X_\nu X_\mu}.
 \end{aligned}
 \label{eq::periodicity}
\end{equation}
Similarly $e^{\mathrm{i}(\varphi+\pi) Z_\nu} = - e^{\mathrm{i}\varphi Z_\nu}$ holds. Hence a shift of $\pi$ in the parameters can only lead to a global phase of $-1$ when one applies the HVA \eqref{eq::HVA_basic} of the TFIM to a state.\\

For the process of optimization, we use the trust-constraint method \cite{Sorensen1982}, where  the symmetric rank-one quasi-Newton method (SR1) ~\cite{Grippo2023} is implemented as Hessian update strategy. For large systems, we apply an implementation of the conjugate gradient method and automatic differentiation in Julia \cite{RevelsLubinPapamarkou2016,mogensen2018optim}. In order to obtain the ground-state energy for a parameter region, we use a strategy we call adiabatic optimization, in which we take the optimized parameters corresponding to $J/h$ and plug it as the initial guess parameters for $J/h-\delta$. Here we assume that we start from a global minimum at the initial value of $J/h$. This ensures that the solution space for optimization remains small hence reducing the number of local minima in the considered landscape. In other words, 
\begin{equation}
    \theta^i_{J/h-\delta} = \theta^f_{J/h},
\end{equation}
where $\theta^{i,(f)}_{J/h}$ are the initial (final) parameters. 
Adiabatic optimization heavily cuts down the number of iterations for each point in the physical parameter space. More importantly, adiabatic optimization helps to avoid barren plateaus. 

\subsection{VQE for periodic clusters}\label{sec:periodic}

It is very important for VQE to have a good initial guess. Periodic transverse field Ising chains are easier to solve with VQE due to the presence of translational invariance. Therefore, these systems can help supply a good initial guess to the rectangular graphs as the solution of the periodic chains is already close to the solution of the open boundary rectangular graphs.  In the following section we investigate the performance of HVA for periodic chains. \\

The one-dimensional TFIM on periodic clusters was investigated with the HVA ansatz \textcolor{black}{by} Wiersema \textit{et al.}~\cite{Wiersema2020}. Due to the presence of translational invariance, the HVA uses the same parameter for each sub-layer in Eq.~\eqref{eq::HVA_basic}, hence using only two parameters per layer. In \cite{Wiersema2020} it was found that for a parameter range of $h\in [0.5,1.5]$ around the critical point, $N/2$ layers are sufficient to express the true ground state of the model. Even the limiting cases are not always easy to deal with. While the ground state in the limit $J=0$ can always be expressed by setting all parameters to zero, the other limit, $h=0$, requires a finite number of layers. In \cite{Mbeng2019} it was proved that $N/2$ layers are necessary and sufficient to express the even parity ground state 
\begin{equation}
    \Psi_{\mathrm{gs},\,h=0} = (\ket{\leftarrow \dots \leftarrow} + \ket{\rightarrow \dots \rightarrow})/\sqrt{2}.
\end{equation}
So the same number of layers as needed for the exact solution is also required to express $\Psi_{\mathrm{gs},\,h=0}$. For odd periodic chains, we found that setting all parameters to $\pi/4$ yields $\Psi_{\mathrm{gs},\,h=0}$. For even number of spins, the parameters change with the number of spins.\\

\begin{figure}[t]
    \centering
    \includegraphics[width=80mm, height=60mm]{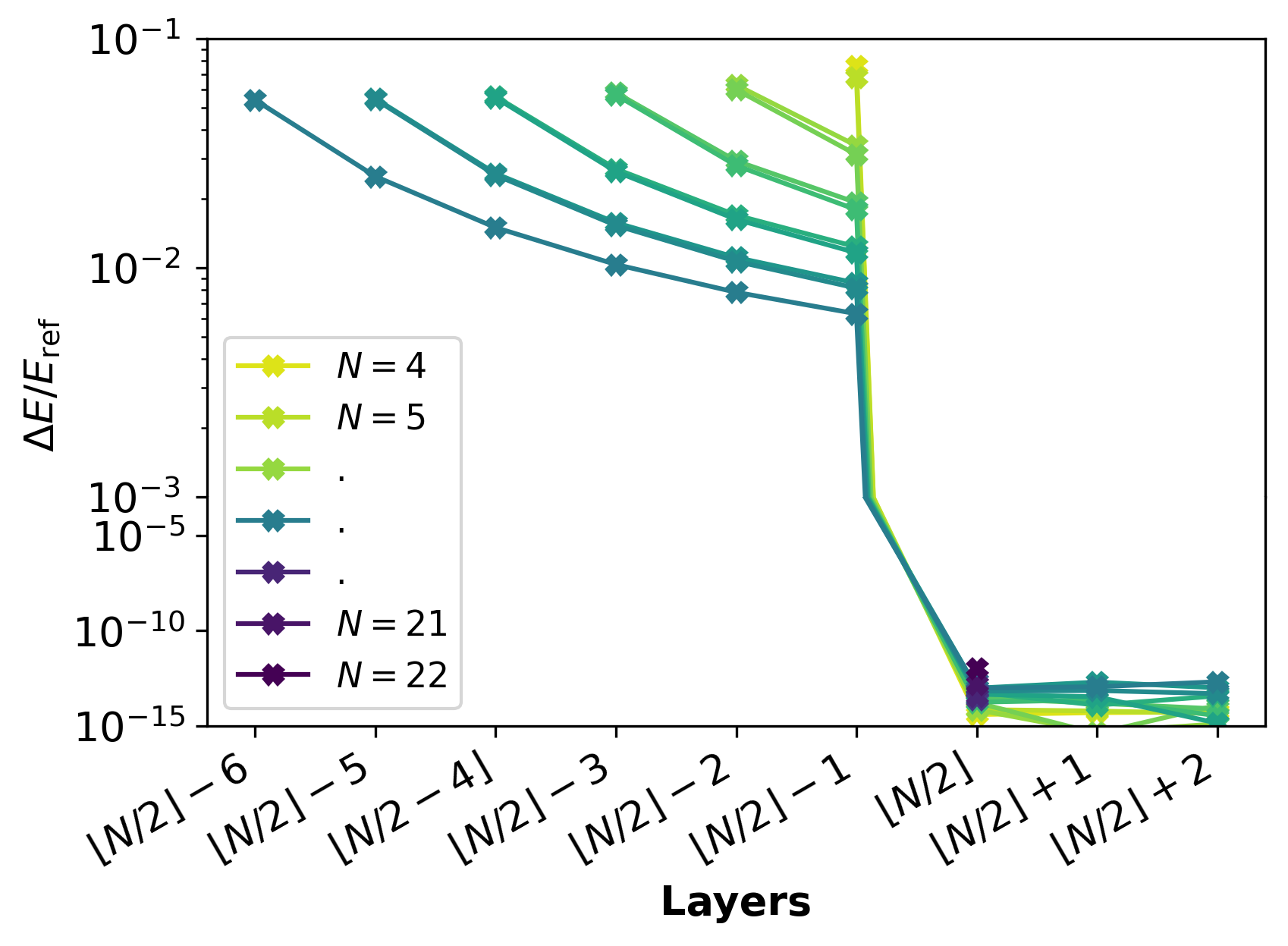}
    \caption{$\Delta E/ E_{\mathrm{\rm ref}}$, the relative difference between the ground-state energy per site calculated using the VQE algorithm ($E_{\rm VQE}$) and that computed using exact diagonalization ($E_{\rm ED} \equiv E_{\rm ref}$ ), is shown for the periodic one-dimensional TFIM. $E_{\mathrm{VQE}}$ is computed using $1$ to $ \lfloor N/2 \rfloor +2  $ number of layers of HVA for $N=$3 to 14 spins. Due to limited computational resources, we calculate the ground state energy for $N=$ 15 to 22 spins at $\lfloor N/2 \rfloor$ layers only.} 
    \label{periodic}
\end{figure}

We reconfirmed the sufficiency and necessity of $\lfloor N/2 \rfloor $ layers in Fig.~\ref{periodic}. It shows the comparison of ground-state energy calculated using the VQE algorithm with that calculated using ED for the ferromagnetic transverse-field Ising chain with periodic boundary conditions at the critical point $J/h=1$. The different curves in the figure show the usage of $1$ to $ \lfloor N/2 \rfloor +2  $ number of layers of HVA for the VQE simulation. The most apparent observation from Fig.~\ref{periodic} is the sudden decrease in the ground-state energy with application of $\lfloor N/2 \rfloor$ layers as compared to lower number of layers. Periodic chains reach accurate ground-state energy at $\lfloor N/2 \rfloor$ layers for a system with $N$ spins. This implies that the VQE simulation for periodic chains requires only $N$ number of parameters and circuit depth of $2N$ following the brick-wall structure of gates. We observe that this also holds true for the whole polarized phase. To sum up, for the TFIM in one dimension, which is exactly solvable, the ground state is not only approximated but really recovered as soon as the opposing limits $J=0$ and $h=0$ can be expressed by the HVA ansatz. Thinking of other models with two phases, this suggests that this is a good requirement for potential layer ansätze with existing perturbative limits.\\

\subsection{VQE as cluster solver for graphs}\label{sec:cluster-solver}


NLCE$+$VQE method requires the calculation of ground-state energy for graphs with open boundary conditions. For the purpose of NLCE$+$VQE, we design a new ansatz motivated by HVA with an additional term for periodic boundaries. We call this new ansatz the periodic Hamiltonian variational ansatz (pHVA). For the TFIM given in Eq.~\ref{ham_tfim}, this pHVA ansatz is given by
\begin{widetext}
 \begin{equation}\label{HVAeqn}
     U_{N_l}^G(\vec{\Theta},\vec{\Theta}_{\mathrm{b}}) = \prod_{l=1}^{N_l}\left [ e^{\mathrm{i}\sum_{\nu=1}^N\theta^{Z}_{\nu,l} Z_\nu}
     \cdot e^{\mathrm{i}\sum_{<\nu,\mu>}\theta^{XX}_{\nu,\mu,l}X_\nu X_{\mu}} e^{\mathrm{i}\sum_{<\nu,\mu>,\mathrm{b}}\theta^{XX}_{\nu,\mu,l,\mathrm{b}}X_\nu X_{\mu}} \right],
 \end{equation}
\end{widetext}
where $\vec{\Theta}_{\mathrm{b}}$ is the set of parameters for the periodic boundaries of the system. These periodic links in the HVA account for the additional parametrized links on the edges of system to establish the periodic boundary in a system for the ansatz. These help to improve the performance of our ansatz for the TFIM with open boundary conditions. Eq.~\ref{HVAeqn} boils down to the conventional HVA for $\vec{\Theta}_{\mathrm{b}}=0$. The pHVA ansatz assumes that the solution of the periodic and the open system are adiabatically connected. The periodic system is translationally invariant and a solution with VQE is found rather easy. In contrast, the open system has extensively more optimization parameters and finding the global minimum is a much harder problem. The pHVA allows to use the knowledge of an already found solution of the periodic system to make the optimization problem of the open system significantly more efficient and, even more important, stable. Apparent from Eq.~\eqref{eq::periodicity}, the pHVA ansatz for the TFIM is periodic in $\pi$ for every component of the vectors $\vec{\Theta}$ and $\vec{\Theta}_{\mathrm{b}}$.\\

\begin{figure}
    \centering
    \includegraphics[height=60mm, width=80mm]{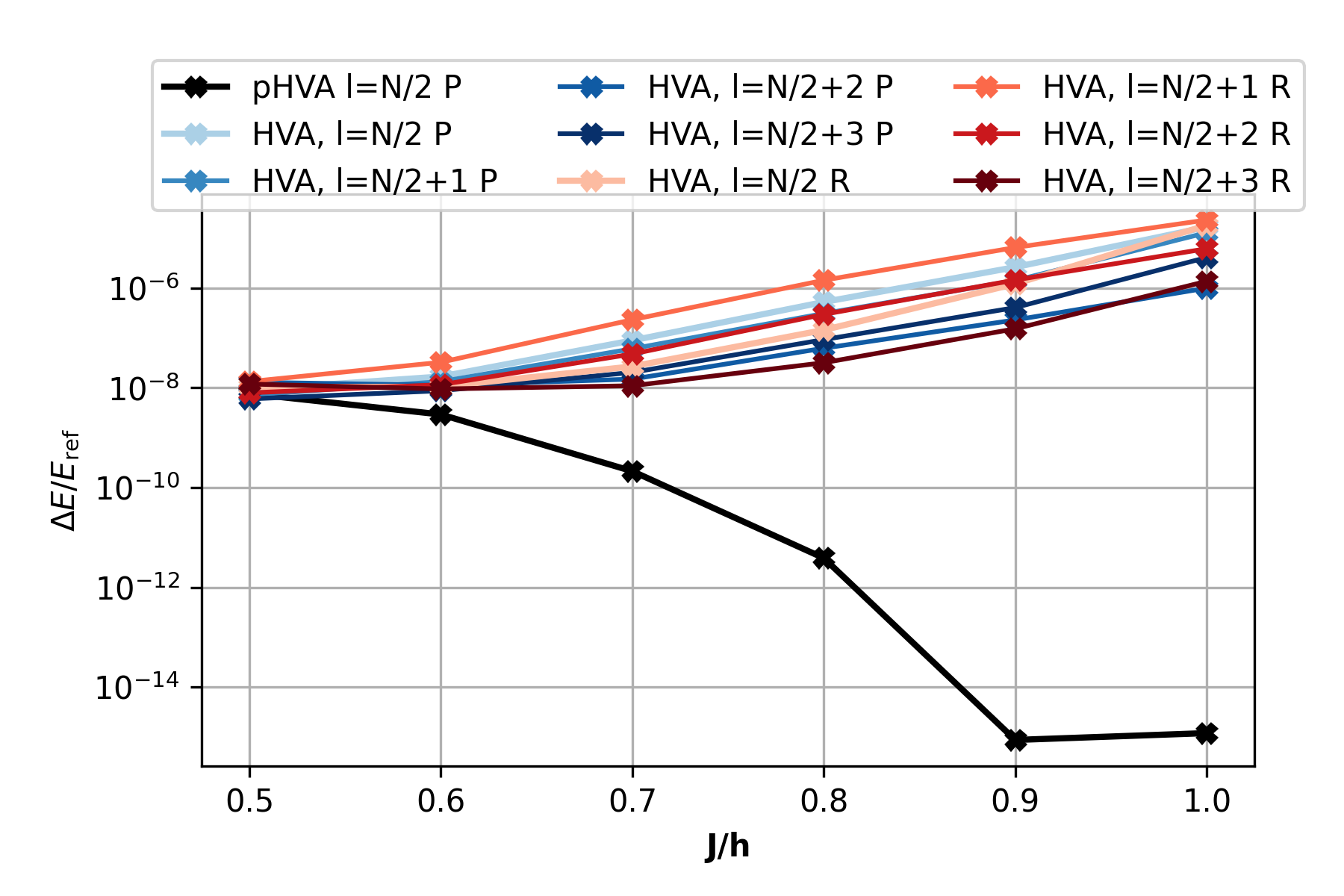}
    \caption{The figure represents the relative error $\Delta E/ E_{\mathrm{ref}}$ in $N=12$ spin open chain segment VQE calculation with respect to exact diagonalization  ($E_{\rm ref} \equiv E_{\rm ED}$) for pHVA  and the traditional HVA. Here `l' is the number of layers used. Letter `P' represents the initial guess used from the periodic chain solution and `R' represents a random initial guess.}
    \label{fig:pHVA-imp}
\end{figure}

 In Fig.\ref{fig:pHVA-imp}, we compare the error in the ground-state energy of $N=12$ chain segment calculated with pHVA and that calculated with HVA using different number of layers up to $l=N/2+3$. It is clearly seen that the accuracy of pHVA is much better ($\approx 10^{-14}$) at the critical point for $N/2$ layers as compared to the conventional HVA with even higher number of layers. We use both random initial guess (represented by `R') as well as the initial guess from the periodic system (denoted by `P'). Nonetheless, in every case, pHVA performs better at capturing the ground state.

\subsection{VQE strategies for finding ground-state energies on graphs}

The TFIM on rectangular graphs does not possess translational invariance like the periodic clusters in one dimension (section ~\ref{sec:periodic}). In contrast to these periodic chains, where the number of parameters for VQE optimization scale as $\mathcal{O}(N)$, the rectangular graphs which have open boundaries in one and two dimensions have of the order $\mathcal{O}(N^2)$ parameters. Gradient-based optimization routines become more and more difficult with increasing number of parameters as the optimization gets stuck in local minima more often and more calculations are needed to calculate Hessians and gradients.\\

However, rectangular graphs in one and two dimensions still possess discrete symmetries. Open transverse-field Ising chains have reflection symmetry. For rectangular graphs in two dimensions the amount of symmetries depends on the graph. These symmetries can be exploited to reduce the number of parameters for VQE simulation to half in one dimension and to at least half in two dimensions. Using these symmetries stabilizes and speeds up the optimization process. \\

In one dimension, starting the adiabatic optimization from $h=0$ with $\Psi_{\mathrm{gs},\,h=0}$ towards larger values of $h$ works much better than starting from $J=0$ ($\ket{\mathrm{ref}}$ is the polarized state). This way a lot of parameters are already fixed to a good value since it is not trivial to express $\Psi_{\mathrm{gs},\,h=0}$. Although not trivial for open chain segments of even length, we still can find out with moderate effort how to choose the unitaries to reach this state as explained in subsection \ref{sec:periodic}. However, for rectangular graphs in general, this is different and we do not know the parameters to express $\Psi_{\mathrm{gs},\,h=0}$. For this reason, we modified our strategy and use an approximation to the ground state at the critical point as starting point for our optimization. We obtain this approximation by taking the solution, i.e. the optimized parameters of the periodic systems. We then map these parameters to the initial guess parameters for the optimization of rectangular graphs. Before describing the mapping in more detail, let us briefly comment what to do in case the critical point is not yet known. For many systems, upper bounds are easily obtained. For example, the TFIM on hypercubic lattices has the largest value of the critical point at $J/h=1$ in one dimension. Hence, in higher dimensions, the critical point would have the range $[0,1]$. In other situations, the adiabatic optimization can always be performed in the direction of lower or higher ratios of $J/h$ starting from some intermediate coupling ratio.\\

For the one-dimensional transverse-field Ising chain, it is possible to directly map the parameters from periodic to open chain segments. Due to the absence of translational invariance, the open chain segments have $N$ parameters per layer after using the reflection symmetry. We map the optimized parameter for periodic chain corresponding to  the $\sum_i X_i X_{i+1}$ term to set it as the initial guess for all the parameters in open chain graphs for $X_i X_{i+1}$ for every layer. The same is done for the $Z_i$ parameters.\\

For the square lattice, the corresponding mapping is not straightforward. Different graphs in rectangular-graph expansion have different mapping schemes for initial guesses from periodic to open boundary graphs. For chain graphs, we use the aforementioned strategy. Concerning other graphs, the outer boundary of rectangular graphs is a finite periodic chain. We initialize the parameters corresponding to the outer boundary by the optimized parameters of the periodic chains. The initial values for the edges and vertices inside the closed loop in case of rectangular graphs are set to zero.\\

\section{Calculation of ground-state energy in the thermodynamic limit}\label{sec:calculation thermodynamic limit}

\begin{figure}[t]
    \includegraphics[height=60mm, width=80mm]{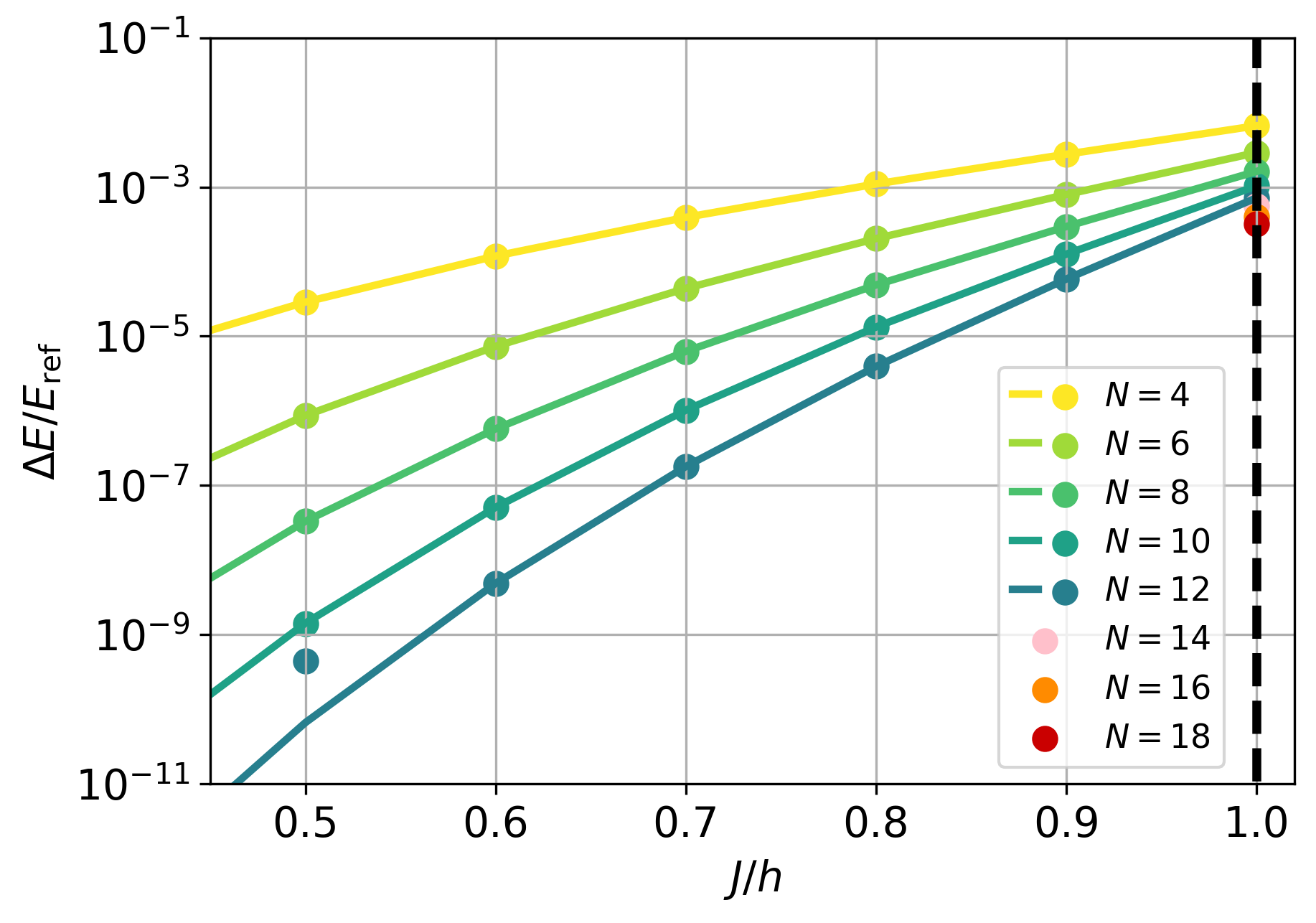}\hfill
    \caption{The figure shows the relative difference between the exact ground-state energy $E_{\rm ref}$ of the one-dimensional TFIM \cite{Pfeuty1970} and that calculated by NLCE$+$VQE (circles) and NLCE$+$ED (solid lines). $N$ represents different largest cluster sizes included in the NLCE calculation. For NLCE$+$VQE we have used $\lceil N/2\rceil$ layers. The vertical dashed line refers to the quantum critical point $J/h=1$. For larger spins (N=14 to 18), we only calculate the data at the critical point due to computational constraints.}
    \label{fig:allspin1d}
\end{figure}

\begin{figure}
    \centering
    \includegraphics[width=80mm]{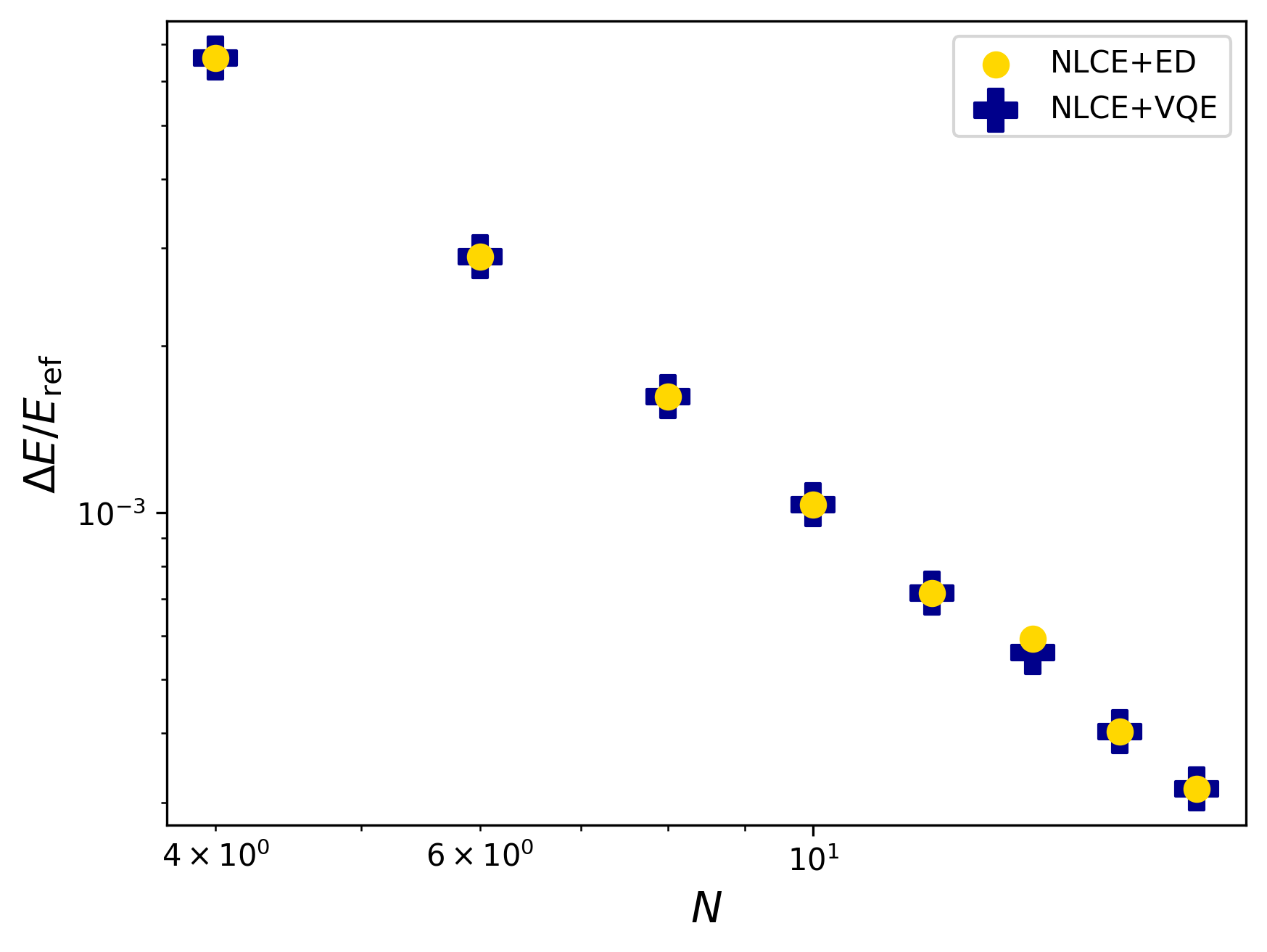}
    \caption{This figure shows the relative difference of NLCE+VQE and NLCE+ED ($\Delta E/ E_{\mathrm{ref}}$) with respect to exact solution $E_{\rm ref}$ plotted against $N$, which is the largest size of the cluster used for the NLCE calculation in the transverse-field Ising chain,  in a log-log plot at the critical point ($J/h=1$) . We perform calculations up to $N=18$ spins. The figure clearly shows that the scaling at the critical point is algebraic.}
    \label{fig:critical_pt_1d}
\end{figure}

\begin{figure}[t]
    \includegraphics[height=60mm, width=80mm]{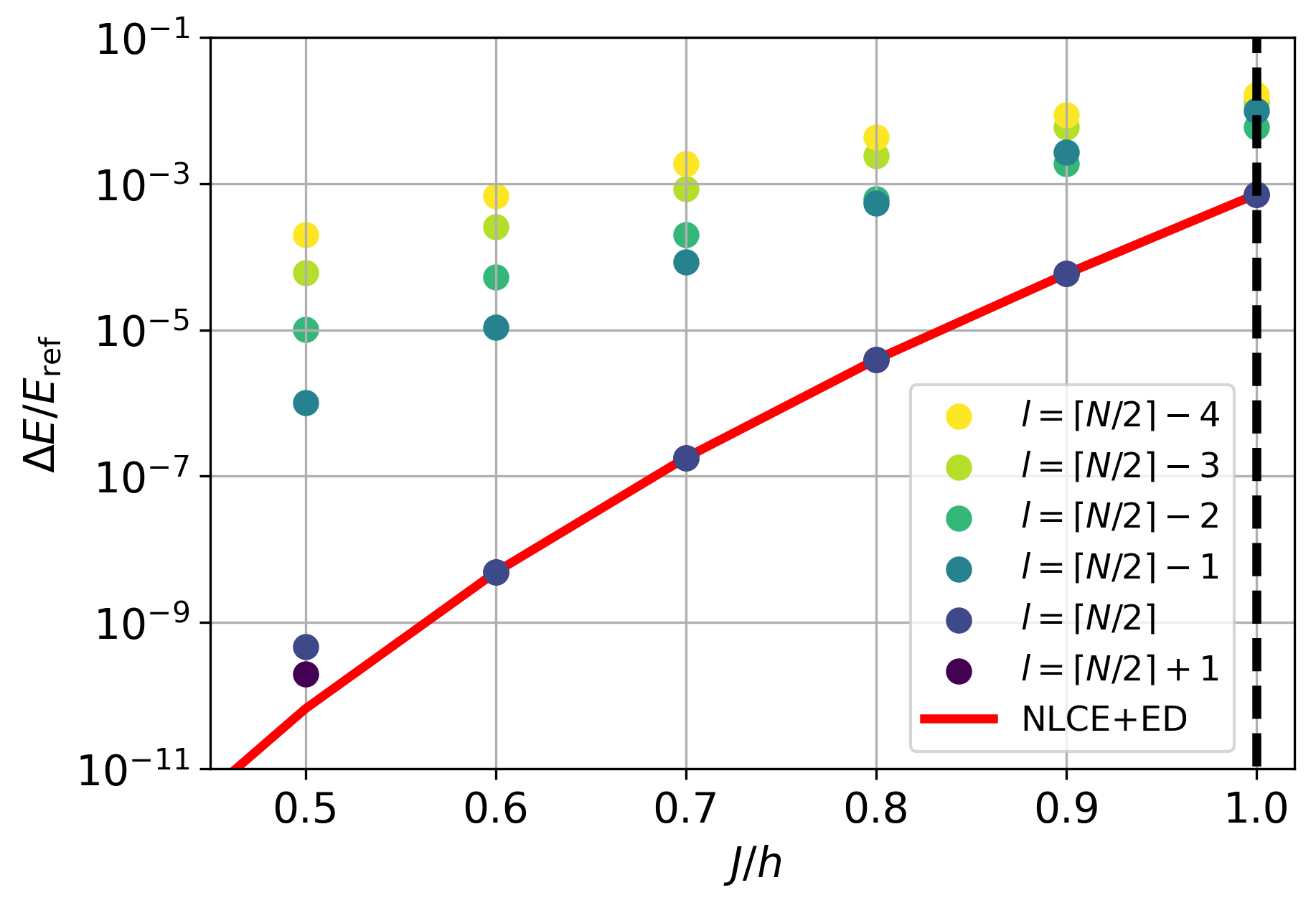}\hfill
    \caption{Performance of NLCE$+$VQE for different number of layers $l$ of pHVA for maximal cluster size of 12 spins for the one-dimensional TFIM. The ground-state energy per site is compared against the exact solution $E_{\rm ref}$ \cite{Pfeuty1970} through relative error shown on the Y axis. The vertical dashed line refers to the quantum critical point $J/h=1$.}
    \label{1d-layers}
\end{figure}

We use NLCE+VQE to approximate the ground-state energy of the TFIM in the polarized phase. We will present results for the model on the chain and on the square lattice using a rectangular graph expansions.\\

\subsection{Results for one-dimensional chain}

For the one-dimensional TFIM we can always use the exact solution \cite{Pfeuty1970} to judge the quality of NLCE$+$VQE. In Fig.~\ref{fig:allspin1d}, we therefore show the difference $\Delta E/ E_{\mathrm{ref}} = e_{\mathrm{NLCE}}/e_{\mathrm{exact}}-1$ between the ground-state energy per site $e_{\mathrm{NLCE}}$ from NLCE and the exact ground-state energy per site $e_{\mathrm{exact}}$ of the transverse-field Ising chain. The different curves represent the size of the largest cluster $N$ used for the NLCE calculation. The accuracy of NLCE increases as the size of the largest cluster increases, i.e. the ground-state energy per site gets closer to the exact result in the thermodynamic limit for increasing $N$. This is true for the whole polarized phase up to the quantum critical point at $J=h$. At the critical point, the convergence of NLCE is slowest compared to $J/h<1$ which can be directly traced back to the vanishing excitation gap yielding an infinite correlation length. The errors of NLCE+VQE and NLCE+ED as compared to the exact solution are shown in Fig. ~\ref{fig:critical_pt_1d}. The error in NLCE$+$VQE in comparison to the exact solution is $\Delta\, e= 9\cdot 10^{-4}$ at $J=h$ using $N=12$. We also performed NLCE+VQE calculation taking $N=14,16$ and $18$ spins as the largest clusters and found the error to be $7\cdot 10^{-4}, 5\cdot 10^{-4}$, and $4 \cdot 10^{-4}$ respectively at the critical point. We therefore can conclude that $\lceil N/2 \rceil$ layers are sufficient to get quantitative results for the transverse-field Ising chain using NLCE$+$VQE. \\

\begin{figure}[t]
    \centering
    \includegraphics[height=60mm, width=80mm]{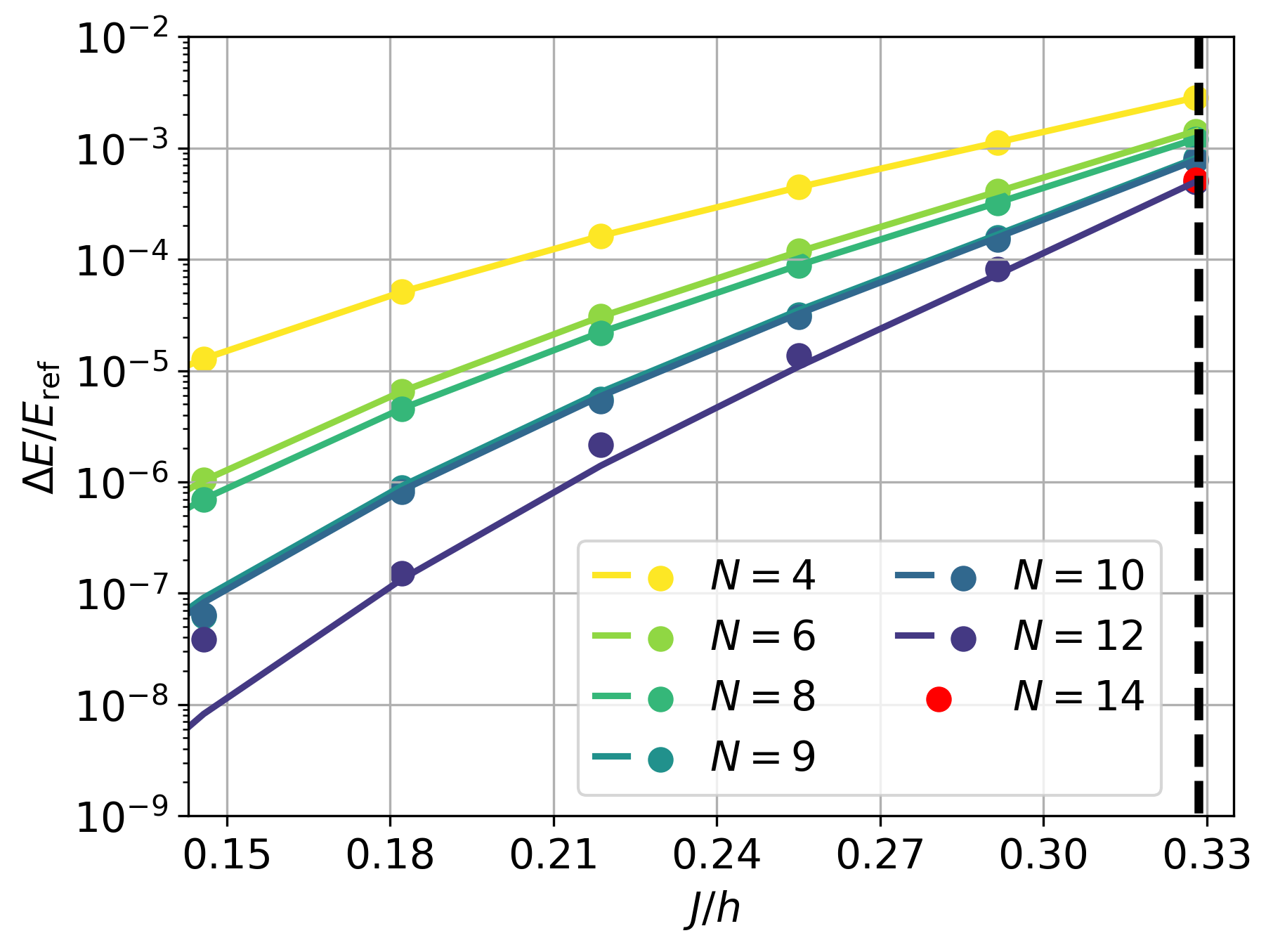}
    \caption{The relative difference $\Delta E/ E_{\mathrm{ref}}$ between the ground-state energy calculated with NLCE$+$ED (solid lines) or NLCE$+$VQE (circles) for different maximal number of sites $N$ and $E_{\rm ref}$ obtained by high-order series expansions about the high-field limit for the TFIM on the square lattice. The NLCE$+$VQE calculation includes all rectangular graphs up to 12 spins. The series have been extrapolated with Pad\'e extrapolations \cite{domb1989phase}. The vertical dashed line refers to the quantum critical point $J/h\approx 0.328$ \cite{He1990, Hesselmann2016}. We also perform NLCE+VQE calculations for $N=14$ at the critical point.}
    \label{fig:square_nlce}
\end{figure}

\begin{figure}
    \centering
    \includegraphics[width=80mm]{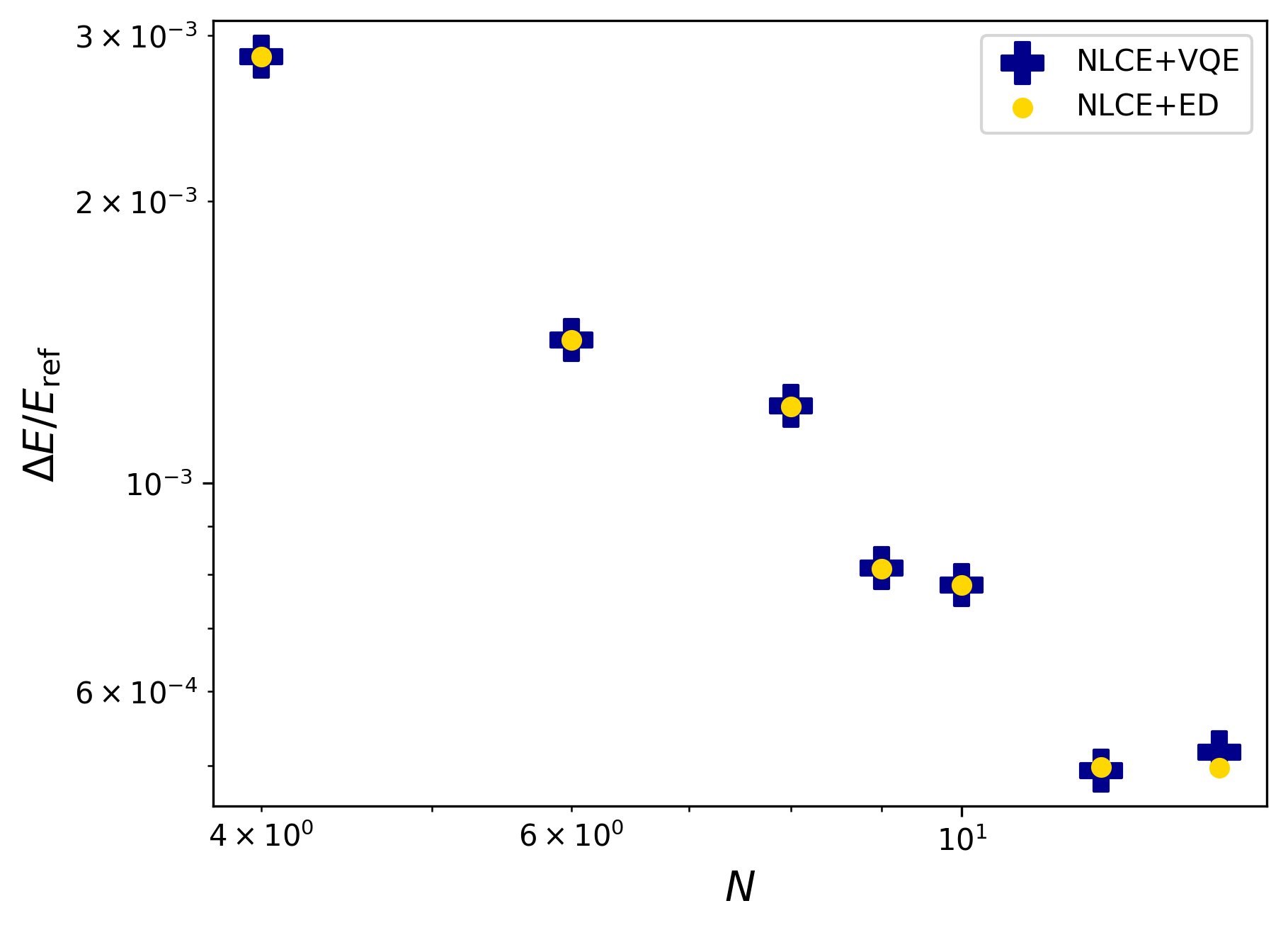}
    \caption{This figure shows the relative difference of NLCE+VQE and NLCE+ED ($\Delta E/ E_{\mathrm{ref}}$) with series expansion $E_{\rm ref}$ plotted against $N$ in a log-log plot, where $N$ is the largest size of the cluster used for NLCE calculation, for the two-dimensional square lattice at the critical point \mbox{$J/h=0.328$}. We perform calculations up to $N=14$ spins.}
    \label{fig:critical_pt_2d}
\end{figure}

\begin{figure}[t]
    \centering
    \includegraphics[height=60mm, width=80mm]{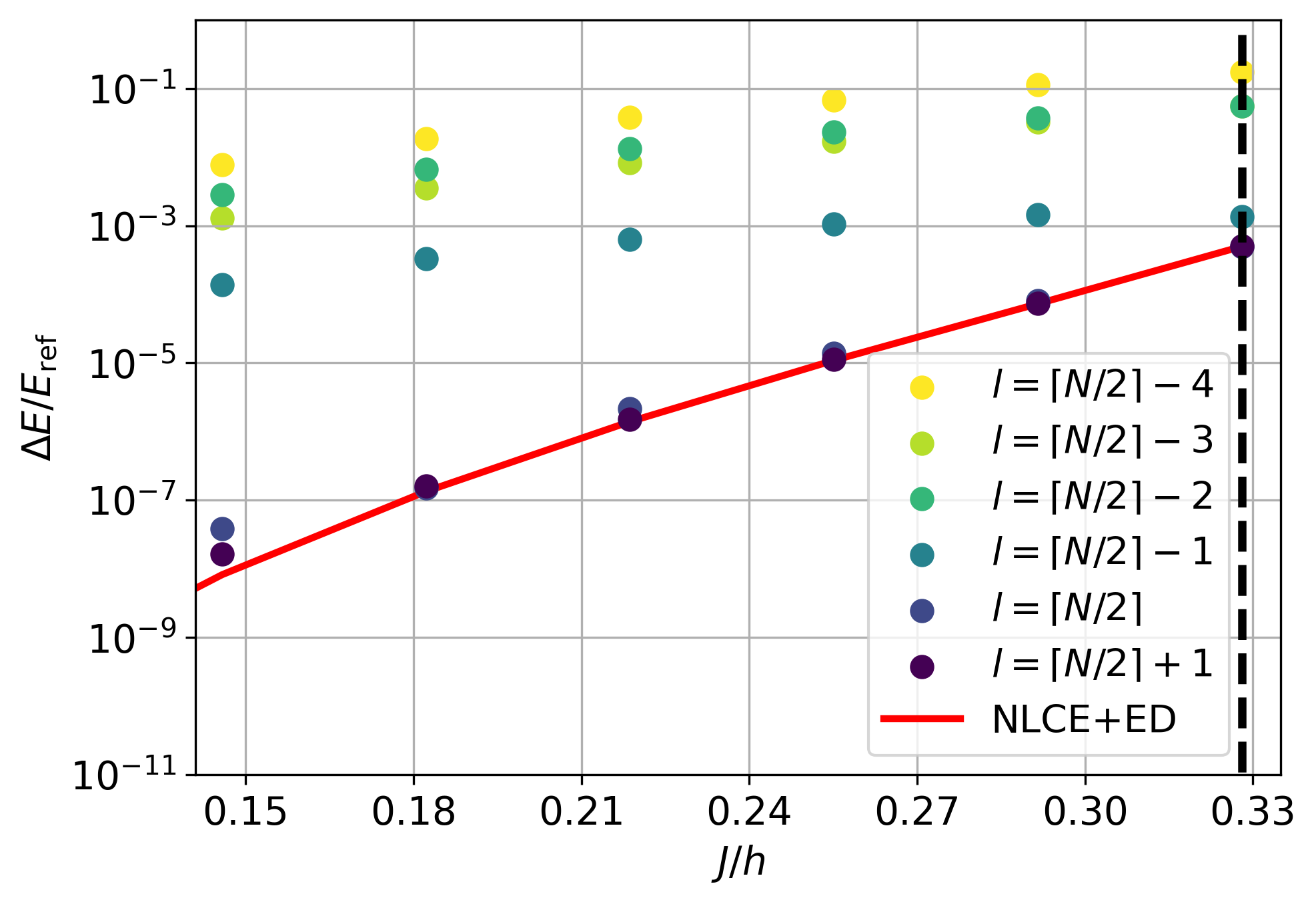}
    \caption{ Error in NLCE$+$VQE for a given number of layers $l$ using pHVA for finite clusters with the largest cluster consisting of 12 spins for the TFIM on the two-dimensional square lattice. $\Delta E/ E_{\mathrm{ref}}$ is the relative difference between the energy of NLCE$+$VQE and $E_{\rm ref}$ from high-order series expansions as described in Fig.~\ref{fig:square_nlce}. The vertical dashed line refers to the quantum critical point $J/h\approx 0.328$ \cite{He1990, Hesselmann2016}.}
    \label{fig:square_layer}
\end{figure}

Next, we discuss the performance of NLCE$+$VQE with the number of layers $l$. The corresponding data for $N=12$ is shown in Fig.~\ref{1d-layers}. Interestingly, the convergence with the number of layers is not smooth, but it rather jumps to larger $\Delta\, e$ when going from $l=\lceil N/2 \rceil$ to $l=\lceil N/2 \rceil-1$. For example, at the critical point $J=h$, one has $\Delta\, e= 9\cdot 10^{-4}$ for $l=\lceil N/2 \rceil$, $\Delta\, e= 1\cdot 10^{-2}$ for $l=\lceil N/2 \rceil-1$, and $\Delta\, e= 7\cdot 10^{-3}$ for $l=\lceil N/2 \rceil-2$. We emphasize that for $l< \lceil N/2 \rceil$  the convergence is not even monotonous, but $\Delta\, e$ is larger for $\lceil N/2 \rceil-1$ layers compared to $\lceil N/2 \rceil-2$ layers. In contrast, increasing the number of layers to $l>\lceil N/2 \rceil$ does not yield any significant further improvement ($\Delta\, e= 9\cdot 10^{-4}$ for $l=\lceil N/2 \rceil+1$). Altogether, this implies that it is necessary to take $\lceil N/2 \rceil$ layers in order to get convergent results for the one-dimensional TFIM. Physically, it is natural to assume that this can be linked to the numerical finding that the exact ground state of periodic TFIM clusters in one dimension can be described by this number of layers. Note that this was indeed the motivation for the pHVA ansatz.\\

\subsection{Results for two-dimensional square lattice}

Let us now turn to the two-dimensional case, where no exact solution is available anymore. We therefore take the ground-state energy from high-order series expansion about the high-field limit \cite{He1990} as a reference for the full polarized phase up to the quantum critical point $J/h\approx 0.328$ \cite{He1990, Hesselmann2016}. More specifically, we use Pad\'e extrapolants of order 16, i.e. Pad\'e extrapolants with numerator $n$ and denominator $m$ so that $n+m=16$, and take their variance as a measure of the uncertainty. Let us stress that this uncertainty ($10^{-5}$ even at the critical point) is well below the accuracy of the NLCE, since the length scales in the series expansion are much larger. This is true for the whole polarized phase up to the quantum critical point at $J/h=0.328$. In Fig.~\ref{fig:square_nlce}, we show the difference $\Delta E/ E_{\mathrm{ref}} = e_{\mathrm{NLCE}}/e_{\mathrm{exact}}-1$ between the ground-state energy per site $e_{\mathrm{NLCE}}$ from NLCE and the extrapolated ground-state energy per site $e_{\mathrm{SE}}$ from series expansions for the TFIM on the square lattice. We perform the NLCE+VQE calculation for different sizes of the largest cluster ranging from $N=4$ to $N=14$ in the rectangular expansion and using $\lceil N/2 \rceil$  number of layers for each cluster. The accuracy of NLCE increases with $N$, i.e. the ground-state energy per site gets closer to the extrapolated series expansion result in the thermodynamic limit. The error in NLCE+VQE and NLCE+ED as compared to the series expansion at the critical point is shown in Fig.~\ref{fig:critical_pt_2d}. At the critical point, the difference to series expansions in NLCE$+$VQE and NLCE+ED is $\Delta\, e= 5\cdot 10^{-4}$ for both $N=12$ and $N=14$ spin clusters. This also implies that the uncertainty of $e_{\mathrm{SE}}$ from the extrapolations does not impact the current discussion. We therefore find that $\lceil N/2 \rceil$ layers are also sufficient to obtain quantitative NLCE$+$VQE results on the square lattice. \\

Similar to 1d, the accuracy of the NLCE+VQE is not smooth with the number of layers as shown in Fig.~\ref{fig:square_layer}. Interestingly, for small $J/h$, the jump in $\Delta\, e$ as a function of the number of layers is even larger in two dimensions. In contrast, at the critical point $J/h=0.328$, $\Delta\, e$ of $l=\lceil N/2 \rceil-1$ layers is already quite close to the $\Delta\, e$ of $l=\lceil N/2 \rceil$ layers, but still a noteworthy energy difference is present. To give numbers: at the critical point, $\Delta\, e= 5\cdot 10^{-4}$ for $l=\lceil N/2 \rceil$, $\Delta\, e= 8\cdot 10^{-4}$ for $l=\lceil N/2 \rceil-1$, and $\Delta\, e= 5\cdot 10^{-2}$ for $l=\lceil N/2 \rceil-2$. Fig.~\ref{fig:square_layer} shows that it is important to take $\lceil N/2 \rceil$ layers to get convergent results for the polarized phase. Taking even a single layer less in the pHVA ansatz influences the quality of the results significantly. Least deviations are seen at the critical point. \\

Thus, we can summarize that a ballpark of $N/2$ layers within the pHVA ansatz is required for the graphs in the NLCE$+$VQE calculation. This does not impose high demands on quantum computation resources such as the number of gates and circuit depths for the calculation of the ground-state energy of the TFIM in the thermodynamic limit.

\subsection{Shot-noise error analysis}

\begin{figure}
    \centering
    \includegraphics[height=70mm, width=90mm]{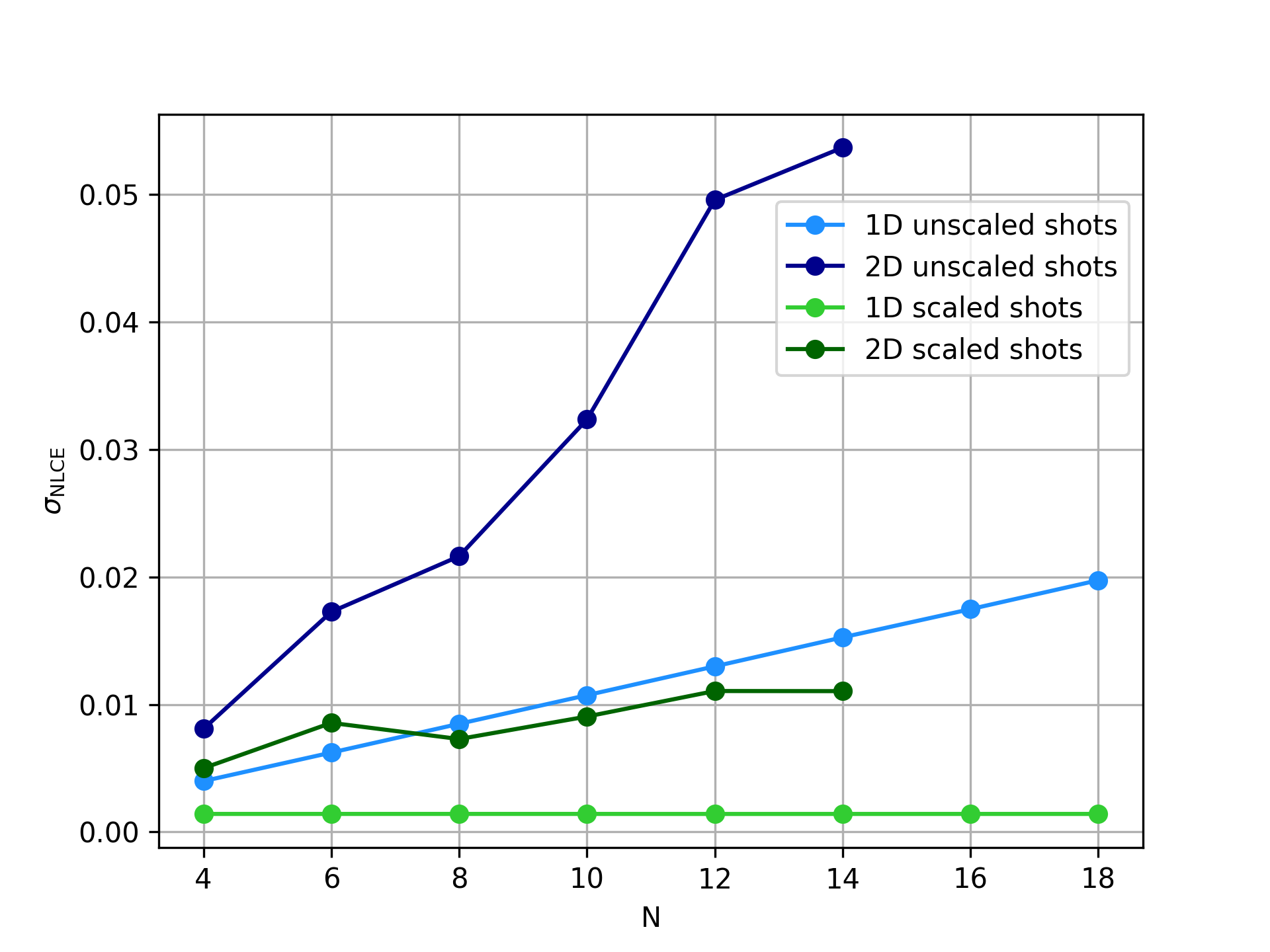}
    \caption{This figure shows the scaling of errors $\sigma_{\rm NLCE}$ in NLCE+VQE with order $N$ of NLCE calculation (i.e. the size of largest cluster used). The standard deviation $\sigma_{\rm NLCE}$ represents the total error in NLCE+VQE due to the noise added to each finite cluster calculation.
    We added random noise from a Gaussian distribution with fixed $\sigma$. We used $\sigma= 10^{-3}$ for the unscaled shots and $\sigma \cdot N$ for the scaled shots.}
    \label{fig:error_N}
\end{figure}

NLCE+VQE is susceptible to error accumulation due to the finite errors in the VQE calculation on each cluster. Apart from optimization errors, other errors arise due to a finite number of measurements on a quantum computer and noise in the devices. As the latter is hard to be defined, we concentrate our analysis here on the shot-noise error and perform simulations to get estimates of the error one could afford in NLCE+VQE calculations. If we add random Gaussian noise of same magnitude on each cluster and estimate the total error of $\sigma_{\rm NLCE}$ of NLCE, we find that the total error depends linearly on the magnitude of the noise. We compare how the total errors scale with the system size in 1d and 2d in Fig.~\ref{fig:error_N}.\\

If we perform a calculation with number of shots that rise linearly with the system size, we expect same errors on each cluster. This is denoted by green curves named as {\it scaled shots} in Fig.~\ref{fig:error_N}, whereas {\it unscaled shots} refers to same number of shots independent of system size. The figure shows different trends in one and two dimensions with better scaling in one dimension and scaled shots. It is observed that the error in NLCE+VQE increases slightly with increasing orders of NLCE but has the same order of magnitude, i.e. $10^{-2}$ for both one- and two-dimensional TFIM up to 14 and 18 spins respectively with added noise of the order $10^{-3}$. For scaled number of shots, the error tends to be constant while for unscaled number of shots, the error increases with the size of the largest cluster used in the NLCE+VQE calculation. \\

We can also determine the scaling of shot-noise rigorously in one and two dimensions. For the chain only two clusters contribute. Hence, to keep the shot-noise error on the same level one needs to increase the number of shots linearly with the system size.\\

For the rectangular expansion in two dimensions, the number of contributions of non-reduced graph energies scales polynomially with $N$. Then in order to keep the same accuracy for the NLCE one needs to increase the number of shots for all graphs with a polynomial of $N$, as there are $\propto N$ graphs contributing to the rectangular expansion. We conclude that the shots only need to be increased polynomially with system size to keep the same error in the rectangular expansion.

\section{Conclusions}\label{sec:conclusion}

In this work, we have introduced NLCE+VQE as an efficient tool to treat quantum many-body systems in the thermodynamic limit combining numerical linked-cluster expansions with the quantum-classical VQE algorithm for the calculations on graphs. Here we aim at quantitative results with our approach. In contrast, conventional VQE studies typically target qualitatively correct ground states of finite systems. Our quantitative approach can be significant in many situations, e.g., when different quantum phases are close in energy like in frustrated systems. We are therefore convinced that {NLCE+VQE} will be useful in the near future for calculations on real quantum-computing devices. Especially, the rather low requirement of only $\lceil N/2 \rceil$ layers of pHVA for the TFIM on clusters of size $N$ is a promising finding. It opens the possibility that the runtime for NLCEs on future quantum computers will scale polynomially with correlation length, whereas NLCEs done with ED on classical computers will always have an exponential scaling. A crucial requirement for this is that the optimization error in the NLCE will also be controlled and gradually decreased with larger system sizes. This is a major obstacle at the moment. In the future we want to approach this problem by trying to use the transformation of smaller clusters as an approximation for the larger ones.\\

This convergence without being completely exact on each graph itself also raises questions regarding improving NLCEs, e.g. whether it is essential to consider exact eigenvalues. For excitation energies, this is not true and leads to a non-convergence of NLCEs in the presence of decay processes \cite{Coester2015}. It is also related to the stability of NLCE+VQE calculations with respect to errors, which are inevitable when doing calculations on current state-of-the-art quantum computers. With respect to that, it would be desirable to construct an approximation for the unitary of larger graphs by using the solutions of smaller subgraphs of this graph. For a conventional NLCE, the logarithm of the transformation $T$, $S=\mathrm{log}\,T$, if properly chosen, is a cluster additive quantity \cite{Hormann2023}, which allows to construct such an approximation. We think it is promising to establish similar considerations for NLCE$+$VQE as it could help to become more robust to errors and potentially avoids problems with local minima. To generalize the concept of cluster additivity to the HVA circuits is thus an important goal for future research. Additionally, if it is easier to adapt this concept for different ansätze than HVA, is a question to be looked into.\\

Summing up, we calculated the ground-state energy for the unfrustrated TFIM on the chain and the square lattice in the thermodynamic limit using NLCE$+$VQE. We reduced optimization errors by introducing pHVA as a modification of HVA, where periodic links are included for calculations on open clusters. Convergence to conventional NLCE results is demonstrated for $\lceil N/2\rceil$ layers. This implies the scaling of the number of layers is proportional to the correlation length $\xi$ in one dimension, since only two chain segments are required for a given $N$. For the square lattice, the situation is more complicated. However, the quadratic scaling $N^2$ of number of rectangular graphs and the linear scaling $l\propto N$ of each cluster are encouraging results. This promises that NLCE$+$VQE is an efficient tool for quantum calculations only having a polynomial scaling of circuit depth with correlation length. To check if this will also hold for other lattice models is an important task for the future.\\

One of our next goals is to implement this method on a real quantum device with measurement sampling and determine its accuracy with acceptable quantum errors. 
To this end, it might be advantageous to treat only the subset of largest possible clusters with VQE and the rest with ED to reduce the overall error of the NLCE.
We also plan to compute various other observables with the VQE method. For example, one can calculate the ground-state expectation value of the static structure factor for the TFIMs calculated in this paper to determine the location of the quantum critical point. Frustrated systems are challenging systems with potential of exhibiting interesting phenomena. Another focus of our research in the future is to explore the viability of NLCE$+$VQE method for frustrated lattice models. Finally, it would be very interesting to generalize the NLCE$+$VQE approach to the calculation of excited states.\\

\section{Acknowledgements}

 The authors thank Michael Hartmann for fruitful discussions on VQE. We also thank Refik Mansuroglu, Lucas Marti and Matthias Mühlhauser for providing their helpful comments and insights. This work was funded by the Deutsche Forschungsgemeinschaft (DFG, German Research Foundation) - Project-ID 429529648 - TRR 306 QuCoLiMa (Quantum Cooperativity of Light and Matter). All authors acknowledge the support by the Munich Quantum Valley, which is supported by the Bavarian state government with funds from the Hightech Agenda Bayern Plus.


%

\end{document}